\documentstyle[11pt,psfig]{article}
\newcommand{\beq}{\begin{equation}}
\newcommand{\eeq}[1]{\label{#1} \end{equation}}
\newcommand{\beqar}{\begin{eqnarray}}
\newcommand{\eeqar}[1]{\label{#1} \end{eqnarray}}
\newcommand{\insertplot}[1]{
\centerline{\psfig{figure={#1},height=11.0cm}}
}
\newcommand{\insertplotshot}[1]{
\centerline{\psfig{figure={#1},height=9.0cm}}
}
\newcommand{\dlt}{\bigtriangleup}
\begin{document}

\begin{center}
{\large {\bf The source of elliptic flow and initial
conditions for hydrodynamical calculations}} \medskip 

{\large
V.K. Magas$^{\star,}$\footnote{Talk given at the New Trends in High-Energy 
Physics, Yalta (Crimea), Ukraine, May 27-June 4, 2000
}, L.P. Csernai,$^{\star,\dagger}$ and Daniel D. Strottman $^\ddagger$} 
\medskip
\end{center}

{\normalsize\noindent
\hspace*{-8pt}$^\star$Section for Theoretical and Computational Physics, 
Department of Physics\\University of
Bergen, Allegaten 55, 5007 Bergen, Norway\\
\hspace*{-8pt}$^\dagger$  KFKI Research Institute for Particle and Nuclear
Physics\\P.O.Box 49, 1525 Budapest, Hungary\\[0.2ex]
\hspace*{-8pt}$^\ddagger$  Theory Division, Los Alamos National Laboratory,\\
Los Alamos, NM, 87454, USA\\[0.2ex]}
\medskip

{\small A model for energy, pressure and flow velocity distributions at the beginning of relativistic heavy
 ion collisions is presented, which can be used as initial condition for hydrodynamical calculations.
The results show that QGP forms a tilted disk, such that the
direction of the largest pressure gradient stays in the reaction 
plane, but
deviates from both the beam and the usual transverse flow
directions. Such  initial
condition may lead to the creation of
"antiflow" or "third flow component" \cite{CR}.
}
\medskip
\paragraph\ {\bf Introduction.}
Fluid dynamical models are widely used to describe ultra-relativistic heavy ion collisions. 
 Their advantage is that one can vary flexibly the Equation of State (EoS) of the matter and 
test its consequences on the reaction dynamics and outcome.  In energetic collisions of 
large heavy ions, especially if Quark-Gluon Plasma (QGP) is formed in the collision, one-fluid 
dynamics is a valid and good description for the intermediate stages of the reaction. Here, 
interactions are strong and frequent, so that other models, (e.g. transport models, string 
models, etc., assuming binary collisions, with free propagation of constituents between collisions) 
have limited validity. On the other hand, the initial and final, Freeze-Out (FO), stages of the reaction
 are outside the domain of applicability of the fluid dynamical model.

In conclusion, the realistic, and detailed description of an energetic heavy ion reaction requires a
 Multi Module Model, where the different stages of the reaction are each described with suitable 
theoretical approaches. It is important that these Modules are coupled to each other correctly: 
on the interface, which is a 3 dimensional hyper-surface in space-time with normal $d\sigma^\mu$, 
all conservation laws should be satisfied (e.g. $[T^{\mu\nu}d\sigma_\nu] = 0$), and entropy should 
not decrease, $[S^\mu d\sigma_\mu] \ge  0$. These matching conditions were worked out and studied
 for the matching at FO in detail in refs.  \cite{Freeze_out}.

The final FO stages of the reaction, after hadronization, can be described well with kinetic models 
where the matter is already dilute.

The initial stages are more problematic. Frequently two or three fluid models are used to remedy 
the difficulties, and to model the process of QGP formation and thermalization. \cite{A78,C82,bsd00} 
Here, the problem is transferred to the determination of drag-, friction- and transfer- terms among the 
fluid components, and a new problem is introduced with the (unjustified) use of EoS in each component 
in a nonequilibrated situations, where EoS does not exist. Strictly speaking this approach can only be 
justified for mixtures of noninteracting ideal gas components.  Similarly, the use of transport theoretical
 approaches assuming dilute gases with binary interactions is questionable, as due to the extreme 
Lorentz contraction, in the C.M. frame, enormous particle and energy densities, with the immediate 
formation of perturbative vacuum should be handled. Even in most parton cascade models these initial 
stages of the dynamics are just assumed in form of some initial condition, with little justification behind.

Our goal in the present work is to construct a model, based on the recent experiences gained in string 
Monte Carlo models and in parton cascades.  One important conclusion of heavy ion research in the last 
decade is that standard 'hadronic' string models fail to describe heavy ion experiments.

All string models had to introduce new, energetic objects:
string ropes \cite{bnk84,S95}, quark clusters \cite{WA96}, fused strings \cite{ABP93},
 in order to describe the abundant formation of massive particles like strange antibaryons. 
Based on this, we describe the initial moments of the reaction in the framework of classical 
(or coherent) Yang-Mills theory, following ref. \cite{GC86} assuming larger field strength 
(string tension) than in ordinary hadron-hadron collisions.  In addition we now satisfy all 
conservation laws exactly, while in ref.  \cite{GC86} infinite projectile energy was assumed, 
and so, overall energy and momentum conservation was irrelevant. We do not solve 
simultaneously the kinetic problem leading to parton equilibration, but assume that the 
arising friction is such that the heavy ion system will be an overdamped oscillator, i.e.
 yo-yoing of the two heavy ions will not occur. This assumption is based on recent string 
and parton cascade results.

\paragraph\ {\bf Formulation of model.}
Our basic idea is to generalize the model developed in \cite{GC86}, 
for collisions of two heavy ions and improve it by strictly satisfying conservation laws.
First of all, we would create a grid in $[x,y]$ plane
($z$ -- is the beam axes, $[z,x]$ -- is reaction plane).
 We will describe the nucleus-nucleus collision in terms of steak-by-streak collisions, 
corresponding to the same transverse coordinates, $\{x_i, y_j\}$.  We assume that baryon 
recoil for both target and projectile arise from the acceleration of partons in an effective field 
$F^{\mu\nu}$, produced in the interaction.  Of course, the physical picture behind this model 
should be based on chromoelectric flux tube or string models, but for our purpose we consider
 $F^{\mu\nu}$ as an effective abelian field. Phenomenological parameters describing this field 
must be fixed from comparison with experimental data.

Let describe the streak-streak collision.
\beq
\partial_\mu \sum_i T_i^{\mu\nu}=\sum_i F_i^{\nu\mu} n_{i \mu} \ ,
\eeq{eq1}
\beq
\partial_\mu \sum_i n_i^\mu = 0 \ , \quad i=1,2\  ,
\eeq{eq2}
$n_i^\mu$ is the baryon current of $i$th nucleus
(we are working in the Center of Rapidity Frame (CRF), which is the same for all streaks. 
The concept of using target and projectile reference frames has no advantage any more).
 We will use the parameterization:
\beq
n_i^\mu=\rho_i u_i^\mu \ ,
\quad
u_i^\mu=(\cosh y_i,\ \sinh y_i)  \ .
\eeq{eq3}
$T^{\mu\nu}$ is a energy-momentum flux tensor. It
consists of five parts, corresponding to both
nuclei and free field energy (also divided into two parts) and one defines the QGP perturbative vacuum.
\beq
T^{\mu\nu}=\sum_i T_i^{\mu\nu}+T^{\mu\nu}_{pert}=
\sum_i\left[ e_i\left(\left(1+c_0^2\right)u_i^\mu u_i^\nu
- c_0^2g^{\mu\nu}\right)
+T_{F,i}^{\mu\nu}\right]+B g^{\mu\nu}\ ,
\quad i=1,2 \ .
\eeq{eq5}
$B$ -- is the bag constant, the equation of state is $P_i=c_0^2 e_i$, where $e_i$ and $P_i$ 
are energy density and pressure of QGP.

In complete analogy to electro-magnetic field
\beq
F_i^{\mu\nu}=\partial^\nu A_i^{\mu}-\partial^\mu A_i^{\nu}=\left(
\begin{array}{cc}
0 & -\sigma_i \\
\sigma_i & 0
\end{array}\right) \ \ ,
\eeq{eq6}
\beq
\sigma_i=\partial^3 A_i^{0}-\partial^0 A_i^{3}\ ,
\eeq{eq7}
\beqar
T_{F,i \mu\nu}=-g_{\mu\nu}{\cal L}_{F,i}+\sum_\beta \frac{{\cal L}_{F,i}}
{\partial \left(\partial^\mu A_i^{\beta}\right)}\partial_\nu A_i^{\beta}
\ \ ,
\eeqar{eq8}
\beq
{\cal L}_{F,i} = - \frac{1}{4}F_{i \mu\nu}F_i^{\mu\nu}\ .
\eeq{eq9}

In our case the string tensions, $\sigma_i$, will have the same absolute value $\sigma$ 
and opposite sign (in complete analogy to the usual string with two ends moving in opposite
 directions), and $\sigma_i$ will be constant in the space-time region after string creation 
and before string decay.

To get the analytic solutions of the above equations, we
use light cone variables
\beq
(z,t)\rightarrow(x^+,x^-),\quad x^\pm=t\pm z \ .
\eeq{eq12}
Following \cite{GC86}, we insist that $e_1, y_1, \rho_1, A_1^\mu$ are functions of $x^-$ only and
 $e_2, y_2, \rho_2, A_2^\mu$ depend on $x^+$ only.

In terms of light cone variables:
\beq
n_i^\pm=n_{i, \mp}=\rho_i(u_i^0\pm u_i^3)=\rho_i e^{\pm y_i} \ ,
\eeq{eq13}
\beq
\left(
\begin{array}{cc}
T_i^{++} & T_i^{+-} \\
T_i^{-+} & T_i^{--}
\end{array}\right)= 
\left(
\begin{array}{cc}
h_{i+} e^{2 y_i} & h_{i-} \\
h_{i-} & h_{i+} e^{-2 y_i}
\end{array}\right) + T_{F,i}\ ,
\eeq{eq14}
where
\beq
h_{i+}=(1+c_0^2)e_i\ ,
\quad
h_{i-}=(1-c_0^2)e_i\ .
\eeq{eq16}
\beq
\left(
\begin{array}{cc}
F_i^{++} & F_i^{+-} \\
F_i^{-+} & F_i^{--}
\end{array}\right)= 
\left(
\begin{array}{cc}
0 & 2\sigma_i \\
* 2\sigma_i & 0
\end{array}\right) \ .
\eeq{eq17}
\beq
T_{pert}=\left(
\begin{array}{cc}
0 & 2 B \\
2 B & 0
\end{array}\right) \ .
\eeq{eq17a}

At the time of first touch of two streaks, $t=0$, there is no string tension.
We assume that strings are created, i.e. the sting tension achieves the 
value $\sigma$ at time $t=t_0$,
corresponding to complete penetration of streaks through each other.

\paragraph\ {\bf Conservation laws --- String rope creation.}
In light cone variables eq. (\ref{eq2}) may be rewritten as
\beq
\partial_-n_1^-+\partial_+n_2^+=0 \ .
\eeq{eq19}
So, we have a sum of two terms, depending on different 
independent variables,
and the solution can be found in the following way.
\beq
\begin{array}{ll}
\partial_-n_1^-=a, & \partial_+n_2^+=-a \ , \\
n_1^-=a x^- + (n_1)_0, & n_2^+=-a x^+ + (n_2)_0 \ .
\end{array}
\eeq{eq20}
Since both $n_1^-$ and $n_2^+$ are positive (and also more or less symmetric) 
we can conclude that for
our case $a=0$.

Finally
\beq
n_1^-=\rho_1 e^{-y_1}=\rho_0 e^{y_0} \ , \quad
n_2^+=\rho_2 e^{y_2}=\rho_0 e^{y_0} \ ,
\eeq{eq21}
\beq
\rho_1=\rho_0 e^{y_0+y_1}\ , \quad  \rho_2=\rho_0 e^{y_0-y_2}  \ .
\eeq{eq21a}

Let us come back to the energy-momentum tensor $T^{\mu\nu}$.
Based on eqs. (\ref{eq7}, \ref{eq8}, \ref{eq9}) and taking 
into account the
Lorentz gauge, $\partial^0 A_i^0 - \partial^3 A_i^3 = 0$, we 
can find
\beq
\left(
\begin{array}{cc}
T_{F,i}^{++} & T_{F,i}^{+-} \\
T_{F,i}^{-+} & T_{F,i}^{--}
\end{array}\right)= 
\left(
\begin{array}{cc}
\sigma_i^2 -2\sigma_i \left(\partial^0 A_i^0\right)
+2\sigma_i \left(\partial^3 A_i^0\right) & 0 \\
0 & \sigma_i^2 +2\sigma_i \left(\partial^0 A_i^0\right)
+2\sigma_i \left(\partial^3 A_i^0\right)
\end{array}\right) \ .
\eeq{eq22}
As mentioned before, after string creation,
i.e. $t>t_0$, and before string decay we choose the string 
tensions in the
form:
\beq
\sigma_2=-\sigma_1=\sigma>0\ .
\eeq{eq23}
To satisfy the above choice and the Lorentz gauge condition we take the 
vector potentials in the following form:
\beq
\begin{array}{ll}
A_1^+ = 0, & A_1^- = -2 \sigma x^- \ , \\
A_2^+ = -2 \sigma x^+ ,& A_2^- = 0 \ .
\end{array}
\eeq{eq24}
In our
calculations we used the parameterization:
\beq
\sigma=A\left(\frac{\varepsilon_0}{m}\right)^2\rho_0
\sqrt{l_1l_2} \ ,
\eeq{eq44}
where $l_1,\ l_2$ are the lengths of initial streaks. The typical values of
$A$ are around $0.05-0.06$. Notice, that
there is only one free parameter in parameterization (\ref{eq44}).
The typical values of $\sigma$ are
$8-15\ GeV/fm$ for $\varepsilon_0=100 \ GeV/nucl$.

The problem with eq. (\ref{eq1}) is that we do not know what the really conserved quantities are.
Using the definition of $F^{\mu\nu}$, eq. (\ref{eq7}), we can
rewrite eq. (\ref{eq1}) as
\beq
\partial_\mu T^{\mu\nu} =\sum_i F_i^{\mu\nu}n_{i, \mu}= 
\sum_i\left(\partial^\mu \left(A_i^\nu n_{i, \mu}\right) -
A^\nu \partial^\mu n_{i, \mu}
-\partial^\nu \left(A_i^\mu n_{i, \mu}\right)
+A^\mu \partial^\nu n_{i, \mu}
\right) \ .
\eeq{eq25}
The solution for $n_1^-$ and $n_2^+$, eq. (\ref{eq21}),
shows as that second and fourth terms vanish. So, we can define 
new
energy-momentum tensor $\tilde{T}^{\mu\nu}$, such that
\beq
\partial_\mu \tilde{T}^{\mu\nu}=0 \ ,
\eeq{eq26}
\beq
\tilde{T}^{\mu\nu}=\sum_i \tilde{T_i}^{\mu\nu}+T^{\mu\nu}_{pert}=
\sum_i \left(T_i^{\mu\nu} - A_i^\nu n_i^\mu +
g^{\mu\nu} A_i^\alpha n_{i \alpha} \right)+B g^{\mu\nu}
\eeq{eq27}
Using the exact definition of $A_i^\mu$ -- eqs. (\ref{eq24}) --
we obtain
$$
\tilde{T}^{\mu\nu}=\left(
\begin{array}{cc}
h_{1+}e^{2y_1}+5\sigma^2 & h_{1-} + 4\sigma x^- n_1^+\\
h_{1-} + 2\sigma x^- n_1^+ & h_{1+}e^{-2y_1}+\sigma^2-2\sigma x^- 
n_1^-
\end{array}\right)+\left(
\begin{array}{cc}
h_{2+}e^{2y_2}+\sigma^2+2\sigma x^+ n_2^+ & h_{2-} - 2\sigma x^+ n_2^-\\ h_{2-} - 
4\sigma x^+ n_2^- & h_{2+}e^{-2y_2}+5\sigma^2 \end{array}\right)
$$
\beq
+\left(
\begin{array}{cc}
0 & 2 B \\
2 B & 0
\end{array}\right) \ .
\eeq{eq27a}
Now the new conserved quantities are
\beq
Q_0=\int \tilde{T}^{00} dV =\sum_i \int_{\Omega_i} \tilde{T_i}^{00} dV
\ ,
\eeq{eq28}
\beq
Q_3=\int \tilde{T}^{03} dV =\sum_i \int_{\Omega_i} \tilde{T_i}^{03} dV
\ .
\eeq{eq29}
Based on conservation of $Q_0,\ Q_3$ we can calculate rapidity,
energy and baryon densities at the moment $t=t_0$,
when the string with tension $\sigma$ is created. These new quantities 
are used as initial conditions
for our differential eqs. (\ref{eq1}, \ref{eq2}).
\beq
\left(\frac{\varepsilon_1(t_0)}{m}\right)=
\left(\frac{\varepsilon_0}{m}\right)\frac{1}{1{+}c_0^2}-
\frac{\sigma^2}{\rho_0 \varepsilon_0 (1{+}c_0^2)}
\frac{l_1{+}5l_2}{4l_1}-
\frac{\sigma e^{y_0}}{8\varepsilon_0(1{+}c_0^2)}(l_1{+}2l_2)
-\frac{B}{\rho_0 \varepsilon_0 (1{+}c_0^2)} \frac{l_1{+}l_2}{2l_1} \ ,
\eeq{eq42}
\beq
\left(\frac{\varepsilon_2(t_0)}{m}\right)=
\left(\frac{\varepsilon_0}{m}\right)\frac{1}{1{+}c_0^2}-
\frac{\sigma^2}{\rho_0 \varepsilon_0(1{+}c_0^2)}
\frac{l_2{+}5l_1}{4l_2}-
\frac{\sigma e^{y_0}}{8\varepsilon_0(1{+}c_0^2)}(l_2{+}2l_1)
-\frac{B}{\rho_0 \varepsilon_0 (1{+}c_0^2)} \frac{l_1{+}l_2}{2l_2} \ .
\eeq{eq43}
Here the $\varepsilon_i$ is energy per nucleon in CRF.
Now the proper baryon density can be found, $\rho_i(t_0)=\rho_0 \frac{\varepsilon_0}{\varepsilon_i(t_0)}$, 
$\gamma_i=\frac{1}{\sqrt{1-v_i^2}}=\frac{\varepsilon_i(t_0)}{m}$.

For $x^\pm>x_0$, 
where $x_0=2t_0-|z(0)|$ defines the string creation surface $t=t_0$, 
for nucleon or cell element in the position $z=z(0)$ at the time $t=0$, 
 we should solve eqs. (\ref{eq26}), with boundary conditions
\beq
\begin{array}{ll}
n_1^\pm (x^-=x_0)=\rho_0 e^{\mp y_0} &
n_2^\pm (x^+=x_0)=\rho_0 e^{\pm y_0} \\ & \\
h_{1+}(x^-=x_0)=e_1(t_0)(1+c_0^2) &
h_{2+} (x^+=x_0)=e_1(t_0)(1+c_0^2) \\ & \\
y_1(x^-=x_0)=y_1(t_0) & y_2(x^+=x_0)=y_2(t_0) \\ & \\
\sigma_1(x^-=x_0)=-\sigma & \sigma_2(x^+=x_0)=\sigma\ ,
\end{array}
\eeq{eq45}
where 
$e_i(t_0)=m\rho_i(t_0)$ -- energy density in the rest frame of $ith$ nuclei.

Let us present the complete analytical solution 
in the following form
\beq
e^{(-)^{i+1}2y_i}=-\frac{d_i}{b_i}+
\left(\frac{d_i}{b_i}+e^{(-)^{i+1}2y_i(t_0)}\right)
\left(1-\frac{x^i-x_0}{\tau_i}\right)^{-\frac{b_i}{\alpha a_j}} \ ,
\eeq{eq54}
\beq
h_{i+}=e^{(-)^{i+1}2y_i}e_i(t_0)(1+c_0^2)e^{-(-)^{i+1}2y_i(t_0)}
\left(1-\frac{x^i-x_0}{\tau_i}\right)\ ,
\eeq{eq55}
\beq
\rho_i=\rho_0e^{y_0}e^{(-)^{i+1}y_i} \ ,
\eeq{eq56}
where $x^1=x^-,\ x^2=x^+$, $i,j=1,2\ ,\ i\not=j$, \ \ \
and  using the notations
\beq
b_i=\alpha a_j+2\sigma\rho_0e^{y_0} \ ,
\eeq{eq57.b}
\beq
d_i=c_i-2\sigma\rho_0e^{y_0}e^{(-)^{i+1}2y_i(t_0)} \ ,
\eeq{eq57.d}
\beq
\tau_i=\frac{e_i(t_0)(1+c_0^2)}{e^{(-)^{i+1}2y_i(t_0)}a_j} \ ,
\eeq{eq58}
\beq
a_1=c_1+2\sigma\rho_0e^{y_0}-2\sigma\rho_0 e^{y_0}e^{2y_1(t_0)}
\ , \quad
a_2=c_2+2\sigma\rho_0e^{y_0}-2\sigma\rho_0 e^{y_0}e^{-2y_2(t_0)} \ ,
\eeq{eq52}
\beq
c_i=\alpha ((1+c_0^2)e_i(t_0)-e_0)/2t_0 \ .
\eeq{eq61}

Then the trajectories of nucleons (or cell elements)
for both nuclei are given by:
\beq
\begin{array}{c}
x_1^+(x^-)=z(0)+\int_{t_0}^{x^-}dx\ e^{2y_1(x)}=\nonumber \\ \\ z(0)
-\frac{d_1}{b_1}(x^--x_0)+
\left(\frac{d_1}{b_1}+e^{2y_1(t_0)}\right)
\tau_1\frac{\alpha a_2}{2\sigma\rho_0e^{y_0}}
\left[
\left(1-\frac{x^--x_0}{\tau_1}\right)^{-\frac{2\sigma\rho_0e^{y_0}}
{\alpha a_2}}
- 1\right] \ ,\nonumber
\end{array}
\eeq{eq59}
\beq
\begin{array}{c}
x_2^-(x^+)=-z(0)+\int_{t_0}^{x^+}dx\ e^{-2y_2(x)}=\nonumber \\ \\ -z(0)
-\frac{d_2}{b_2}(x^+-x_0)+\left(\frac{d_2}{b_2}+e^{-2y_2(t_0)}\right)
\tau_2\frac{\alpha a_1}{2\sigma\rho_0e^{y_0}}
\left[
\left(1-\frac{x^+-x_0}{\tau_2}\right)^{-\frac{2\sigma\rho_0e^{y_0}}
{\alpha a_1}}
- 1\right] \ ,\nonumber \end{array} \eeq{eq60}
for nucleon or cell element in the position 
$z=z(0)$ at the time $t=0$.

\paragraph\ {\bf Recreation of the matter.}
As we may see from the trajectories, eqs. (\ref{eq59}, \ref{eq60}), 
nucleons
(or cell domains) will keep going in the initial direction up to the
time $t=t_{i,turn}$, then they will turn and go backwards until the 
two streaks again
penetrate through each other and new oscillation will start. Such a 
motion is
analogous to the "Yo-Yo" motion in the string models. 
Of course, it is difficult to 
believe that
such a process would really happen in heavy ion collisions, because of string decays, 
string-string interactions, interaction between streaks and other reasons, which are quite 
difficult to take into account.  To be realistic we should stop the motion described by  
eqs. (\ref{eq59}, \ref{eq60}) at some moment before the projectile and target cross again.

We assume that the final result of collisions of two streaks 
after stopping the string's expansion and  after its decay,
is one streak with length $\dlt l_f$, homogeneous energy density
distribution, $e_f$, and baryon charge distribution, $\rho_f$, moving
like one object with rapidity $y_f$. We assume that this is due to
string-string interactions and string decays. As it was mentioned above 
the typical values
of the string tension, $\sigma$,
are of the order of $10\ GeV/fm$, and these may be treated as
several parallel strings. The string-string interaction will produce 
a kind of
"string rope" between our two streaks, which is
responsible for final
energy density and baryon charge homogeneous distributions.  
Now it is worth to mention that decay of our "string rope" does not allow charges to remain at the 
ends of the final streak, as it would be if we assume full transparency.

The homogeneous distributions are the simplest assumptions, which may be modified based 
on experimental data. Its advantage is a simple expression for $e_f,\ \rho_f, 
\ y_f$.

The final energy density and rapidity, $e_f$ and $y_f$, may be 
determined
from conservation laws.
\beq
\cosh^2 y_f = \frac{(M^2(1+c_0^2)+2c_0^2)+
\sqrt{(M^2(1+c_0^2)+2c_0^2)^2+4c_0^4(M^2-1)}}{2(1+c_0^2)(M^2-1)} \ ,
\eeq{eq66}
where we neglected $B \dlt l_f$ next to $Q_0$ and introduced the 
notation
$M=(l_2+l_1)/(l_2-l_1)$,
\beq
e_f=\frac{{Q_0\over\dlt x \dlt y} -B\dlt l_f}{((1+c_0^2)\cosh^2 y_f 
- c_0^2)
\dlt l_f} \ .
\eeq{eq68}

\begin{figure}[htb]
\centering\noindent
 \insertplotshot{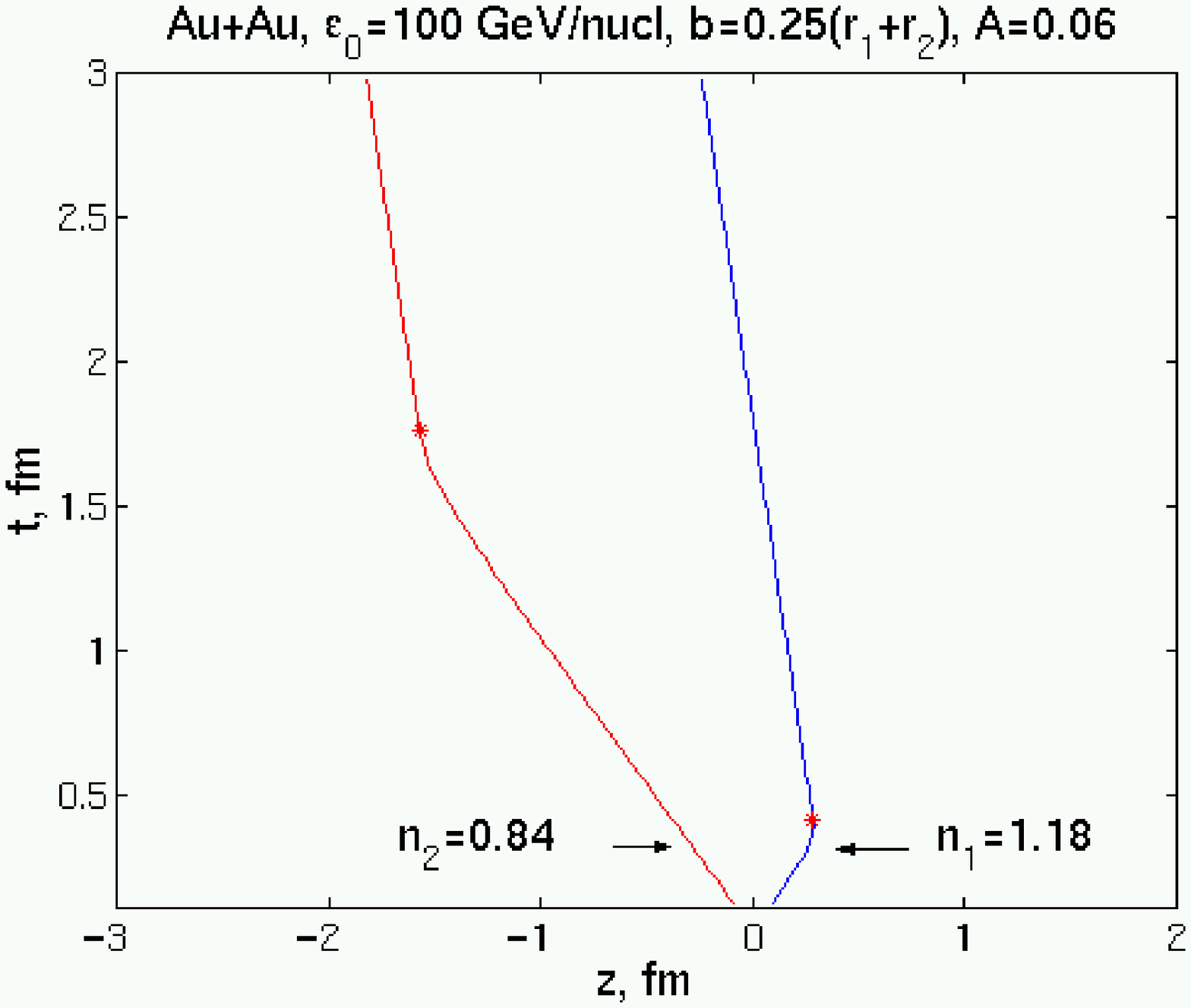} 
\caption{The typical trajectory of the ends of two 
initial streaks, corresponding to numbers of nucleons, $n_1$ and $n_2$. Stars denote the stopping 
and turning points, where $y_i=y_f$. From $t_0$ to the 
turning points streak ends keep going in their 
initial direction according to eqs. (\ref{eq59}, \ref{eq60}). 
Later the final streak starts to move 
like one object with rapidity, $y_f$ (\ref{eq66})
in CRF.}
\label{fig2}
\end{figure}

The typical trajectory of the streak ends is presented in Fig.  \ref{fig2}. From $t_0$ they move 
according to eqs. (\ref{eq59}, \ref{eq60}) until they reach the rapidity $y_i=y_f$.
Later the final string starts to move
like one object with rapidity $y_f$.

The turning points can be found from the condition:
\beq
y_i=y_f \ ,
\eeq{eq62}
which gives for $i$th nucleus ($x_1=x^-,\ x_2=x^+$)
\beq
x_{i,\ turn}=x_0+\tau_i
\left[1-\left(\frac{\frac{d_i}{b_i}+e^{(-)^{i+1}2y_i(t_0)}}
{\frac{d_i}{b_i}+e^{(-)^{i+1}2y_f}}
\right)^{\frac{\alpha a_j}{b_i}}\right] \ .
\eeq{eq63}

\begin{figure}[htb]
\centering\noindent
\insertplot{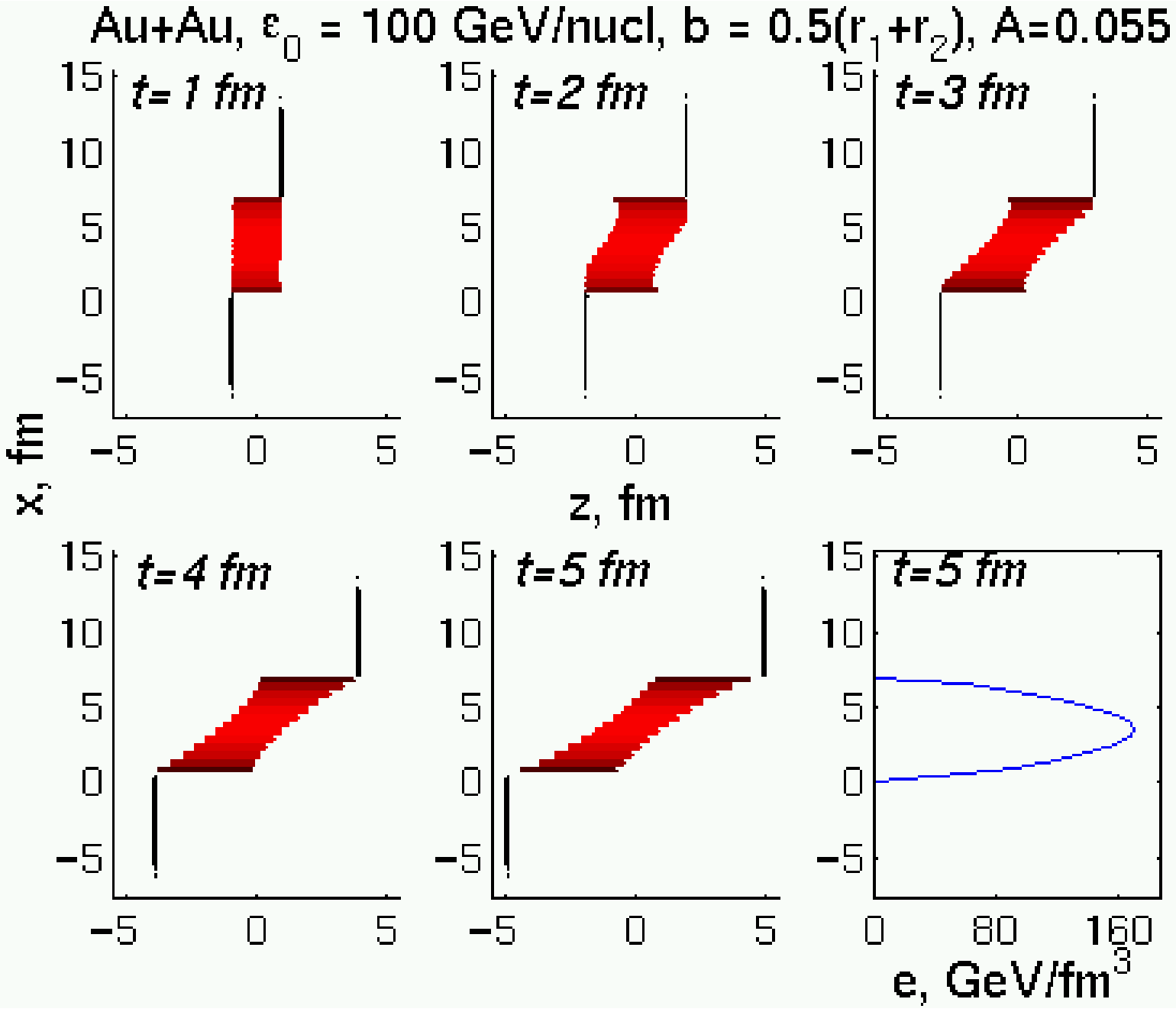}
\caption{The Au+Au collisions, $\varepsilon_0=100\ GeV/nucl$, 
$b=0.5(R_1+R_2)$, $A=0.055$ (parameter $A$ introduced in (\ref{eq44})),
 $y=0$ (ZX plane through 
the centers of nuclei). We would like to notice that 
final shape of QGP volume is a tilted disk $\approx 45^0$, 
and the direction of the fastest expansion will deviate from both 
the beam axis and the usual transverse flow direction, and might be a 
reason for the third flow component, as argued in \cite{CR}.
 }
\label{ev11}
\end{figure}

\paragraph\ {\bf Initial conditions for hydrodynamical calculations.} 
In this section we present the results of our calculations. We are interested in the shape of 
QGP formed, when string expansions stop and their matter is 
locally equilibrated. 
This will be the initial state for further hydrodynamical calculations. We may see in 
Figs. \ref{ev11}, that QGP forms a tilted disk for $b\not =0$. So, 
the direction of fastest 
expansion, the same as largest pressure gradient, 
will be in the reaction plane, 
but will deviate from both the beam axis
and the usual transverse flow direction.  So, the 
new flow component, called "antiflow" or "third flow component", 
will appear in addition to the usual transverse flow component
in the reaction plane. 
With increasing beam energy the usual transverse flow is getting 
weaker, while this new flow component 
is strengthened. The mutual effect of the usual directed transverse flow and this new "antiflow" or 
"third flow component" lead to an enhanced emission in 
the reaction plane.  This was actually 
observed and widely studies earlier 
and referred to as "elliptic flow".

\paragraph\ {\bf Conclusions.}
Based on earlier Coherent Yang-Mills field theoretical models, and introducing effective 
parameters based on Monte-Carlo string cascade and parton cascade model results, a 
simplified model is introduced to describe the pre fluid dynamical stages of heavy ion 
collisions at the highest SPS energies and above.  
The  model  predicts limited transparency for massive heavy ions. 

Contrary to earlier expectations, --- based on standard string tensions of 1 GeV/fm 
which lead to the Bjorken model type of initial state, --- effective string tensions 
are introduced for collisions of massive heavy ions, as a consequence of collective 
effects related to QGP formation.  These collective effects in central and semi central 
collisions lead to an effective string tension of the order of 10 GeV/fm and consequently 
cause much less transparency than earlier estimates. The resulting initial locally 
equilibrated state of matter in semi central collisions takes a rather unusual form, which 
can be then identified by the asymmetry of the caused collective flow.  Our prediction is that 
this special initial state may be the cause of the recently predicted "antiflow" or 
"third flow component". 

Detailed fluid dynamical calculations as well as flow experiments at
semi central impact parameters for massive heavy ions are needed 
at SPS and RHIC energies to connect the predicted special initial state
with observables.

\end{document}